\documentclass[twocolumn,showpacs,amsmath,amssymb]{revtex4}
\usepackage{amsmath} %
\usepackage{graphicx}
\usepackage{amsfonts}% Include figure files
\usepackage{dcolumn}% Align table columns on decimal point
\usepackage{bm}% bold math

\begin{document}

\title{Mobility and density induced amplitude death in metapopulation networks of coupled oscillators}

\author{Chuansheng Shen$^{1,2}$}

\author{Hanshuang Chen$^3$}

\author{Zhonghuai Hou$^1$}\email{hzhlj@ustc.edu.cn}

\affiliation{$^1$Hefei National Laboratory for Physical Sciences at
Microscales \& Department of Chemical Physics, University of
 Science and Technology of China, Hefei, 230026, China \\
$^2$Department of Physics, Anqing Normal University, Anqing, 246011, China \\
$^3$School of Physics and Material Science, Anhui University, Hefei,
230039, China}

\date{\today}

\begin{abstract}
We investigate the effects of mobility and density on the amplitude
death of coupled oscillators in metapopulation networks, wherein
each node represents a subpopulation with any number of mobile
individuals. We perform stochastic simulations of the dynamical
reaction-diffusion processes associated with the Landau-Stuart
oscillators in scale-free networks. Interestingly, we find that,
with increasing the mobility rate or density, the system may undergo
phase transitions from incoherent state to amplitude death, and
then to frequency synchronization. Especially, there exists an extent
of intermediate mobility rate and density leading to global
oscillator death. In addition, we show this nontrivial phenomenon is
robust to different network topologies. Our findings may invoke
further efforts and attentions to explore the underlying mechanism
of collective behaviors in metapopulation coupled systems.
\end{abstract}

%89.75.-k, Complex systems %64.60.aq, Networks
%05.45.-a, Nonlinear dynamics and chaos
%05.45.Xt, Synchronization; coupled oscillators

\maketitle

\section{Introduction}
The complex dynamics of coupled nonlinear oscillators has gained
significant research interest in many branches of science and technology
\cite{PREP2012205}, ranging from the modeling of
biological rhythms in the heart
\cite{JTB19671615,CRE1987704,PhysRevLett.82.3556}, nervous system
\cite{SIAM1990125,CPAM1986623}, intestine \cite{SIAM1984215} and
pancreas\cite{BPJ1988411}. One of the most intriguing emergent
phenomena is the so-called amplitude death (AD), a phenomenon that
coupled  oscillators stop oscillation and
settle to a quenched steady state. Since the
initial work by Bar-Eli \cite{JPC19843616}, AD has been
observed in various real systems, such as the Belousov-Zhabotinsky
reaction system \cite{JPC19892496}, relativistic magnetrons
\cite{PhysRevLett.62.969}, and synthetic genetic networks
\cite{PhysRevLett.99.148103,EPL200928002,Chaos2010023132,PS2011045007}, to list just a few.
The physical mechanism underlying AD has also been extensively investigated\cite{PD1990403,PhysRevLett.65.1701,JSP1990245,PD1991293}.
For coupled oscillators on a regular lattice,
mechanisms leading to AD may include oscillator mismatch
\cite{JPC19843616,PD1990403}, delayed coupling
\cite{PhysRevLett.80.5109,PhysRevLett.85.3381,PhysRevLett.91.094101,
PhysRevE.72.056204,PLA20102636}, asymmetric coupling
\cite{PhysRevE.47.864,PLA2000401,PhysRevE.85.046206}, time-dependent
coupling \cite{PhysRevE.68.067202}, conjugate
coupling \cite{PhysRevE.76.035201,CHAOS2009033143}, nonlinear
(nondiffusive) coupling \cite{PhysRevE.81.027201}, linear
augmentation \cite{PhysRevE.83.067201}, and so on
\cite{PhysRevE.84.046212,PhysRevE.85.046211,PhysRevE.85.057204}. See
a recent review \cite{PREP2012205} for more details, which gives the
deep insight and many clues for understanding AD induced by
different scenarios.

On the other hand, AD in complex networks of coupled oscillators has
also attracted much attention
\cite{PhysRevE.81.027201,PhysRevE.85.056211,PhysRevE.62.6440,PhysRevE.76.016204,
PhysRevE.69.056217,PhysRevE.70.066201,PhysRevE.79.036214,PhysRevE.68.055103,NJP2009093016}.
It has been shown that network topologies, being regular
\cite{PhysRevE.62.6440,PhysRevE.85.056211,PhysRevE.76.016204,PhysRevE.69.056217,PhysRevE.70.066201},
random \cite{PhysRevE.79.036214}, small-world
\cite{PhysRevE.68.055103} or scale-free \cite{NJP2009093016}, may
influence AD in a nontrivial way, by enhancing or suppressing it
with varying network randomness or heterogeneity. Nevertheless,
previous studies on AD in complex networks only deal with the case
of immobile individuals and each network node is occupied by one
single individual. Very recently, the metapopulation network model
\cite{NAP2007276}, which incorporates local node population,
mobility over the nodes,
 and a complex network structure, has drawn intensive attention. This model has been successfully exploited
in different contexts, including epidemic spreading
\cite{PhysRevE.86.036114,JTB2008509,PRE08016111,SREP2011001},
biological pattern formation \cite{SCI20101616,PNAS098429}, chemical
reactions \cite{NAP2010544}, population evolution \cite{JTB201287},
and many other spatially distributed systems
\cite{PR2011001,NAP201232}. It is shown that the density and the
mobility of the individuals could have drastic  impact on the
emergence of collective behaviors in general
\cite{NAP2007276,PRL07148701}, and particularly, density and
mobility induced synchronization of coupled oscillators has been
reported \cite{arXiv:1211.4616,PhysRevE.83.025101}. Therefore, one
may ask: How would the mobility and the density would influence AD
in metapopulation networks of coupled oscillators?

In the present work, we study the emergence of AD in a
metapopulation model which incorporates mobility over a complex
network together with local interactions of the individuals at the
network nodes. We find that, for small mobility rate and lower
density, the metapopulation displays incoherent state, for
intermediate mobility rate or density, all the oscillators
spontaneously cease their oscillations, but for large ones, the
death state can be eliminated and synchronize to a common frequency,
overcoming the disorder in their natural frequencies. Furthermore,
we show this nontrivial phenomenon is robust to different network
topologies.

\section{ Model Description } \label{sec2}

We consider a system of $N$ distinct subpopulations labeled $\mu$,
each corresponding to a network node. The density
$\rho$ of the metapopulation is given by $\rho =
\frac{1}{{{N}}}\sum\nolimits_{\mu = 1}^{{N}} {{N_\mu}}$, where
$N_\mu$ is the number of individuals in node $\mu$.
Individuals inside each subpopulation run stochastically through the
paradigmatic Landau-Stuart oscillators \cite{PhysRevLett.65.1701},
whose dynamics is described by:
\begin{eqnarray}\label{EqModel}
{{\dot z}_j^\mu(t)} & = & {z_j^\mu(t)}(r - |{z_j^\mu}{|^2} +
i{\omega_j^\mu}) +
K(\langle z\rangle^\mu  - {z_j^\mu(t)})
\end{eqnarray}
where $z_j^\mu(t)$ is the complex amplitude of the  ($j=1,..., N_\mu$)
oscillator in the $\mu$-th node,  $r>0$ denotes a homogeneous
oscillation growth rate, $K$ is the coupling
strength, and $\langle z\rangle^\mu =  \frac{1}{{{N_\mu}}}\sum\nolimits_{i =
1}^{{N_\mu}} {{z_i^\mu(t)}}$ is the mean field inside node $\mu$.
 ${\omega _j^\mu}$ denotes the natural frequency of the \emph{j}th oscillator within
 node $\mu$, which is picked up from a certain distribution. In the present work, we adopt
 a linear distribution $ -\gamma \le \omega \le \gamma$ with $\gamma$ a constant.
 Without coupling, the trajectory of each single oscillator will settle
to a limit cycle with frequency $\omega_i^\mu$ and radius $| z_j^\mu(t)|=r$. With sufficiently
strong $K$ and wide distribution of ${\omega_j^\mu}$, the
oscillators pull each other off their limit cycles, and collapse
into a state of zero amplitude $z_j^\mu = 0$, i.e., an AD state.

The above equation actually defines the \lq\lq reaction\rq\rq\,
process that governs the temporal behavior of each individual inside
the metapopulation nodes. We now assume that the individuals can
diffusion randomly among the nodes. The system evolves in time
according to the following rules \cite{NAP2007276}. We introduce a
discrete time step $\tau $ representing the fixed time scale
 of the process. The reaction and diffusion rates are
therefore converted into probabilities. In the reaction step, all
the individuals are updated in parallel according to
Eq.\ref{EqModel}. After that, diffusions take place by allowing each
individual to move into a randomly chosen neighboring node with
probability $D \tau$, where $D$ denotes the mobility rate. If not
otherwise specified, the parameters are $N=1000$, $\tau =0.001$,
$r=0.4$, $\gamma=3.0$, and $K=10$. We choose the mobility rate $D$
and the density $\rho$ as main control parameters. Each simulation
plot is obtained via averaging over 20 independent runs.

\section{Results and Discussion}  \label{sec3}

To begin, we consider
scale-free networks generated by using the Barab\'{a}si--Albert (BA) model
\cite{SCI99000509} with power-law degree distribution $p(k)\sim
k^{-3}$. We fix $\rho =10$ (thus we have totally $N_p=N\rho$ individual oscillators) and vary $D$ to investigate how the
oscillators evolve in time. Initially, the oscillators are homogeneous distributed among the nodes.
If diffusion is absent ($D=0$), each node will stay in an incoherent state for the above-mentioned parameters.
If $D$ is small, each node still remains incoherent, as shown in Fig.1(a) for $D=0.1$, where
typical time series of $Re(z_j^\mu(t))$ for several
randomly-chosen oscillators within a random chosen node are depicted.
For moderate diffusion rate $D=1.0$ as shown in Fig.1(b), the time series
eventually collapse into $\left| {{Z_j^\mu} \approx 0} \right|$,
which indicates the occurrence of global AD. However, for
sufficiently large $D$, say $D=10$ in Fig.1(c), the oscillators are locked to a synchronized state.
Therefore, we observe an interesting mobility induced transition from incoherence to AD and then to synchronization
in our metapopulation oscillator network model.
In addition, we have also perform simulations with fixed $D$ and varying $\rho$. Similar behaviors (time series not shown)
 are found, i,e, the system will also undergo a density induced transition from incoherence to AD and then to
 synchronization with increasing $\rho$ given $D$ is not too small.

\begin{figure}[h]
\centerline{\includegraphics*[width=1.0 \columnwidth]{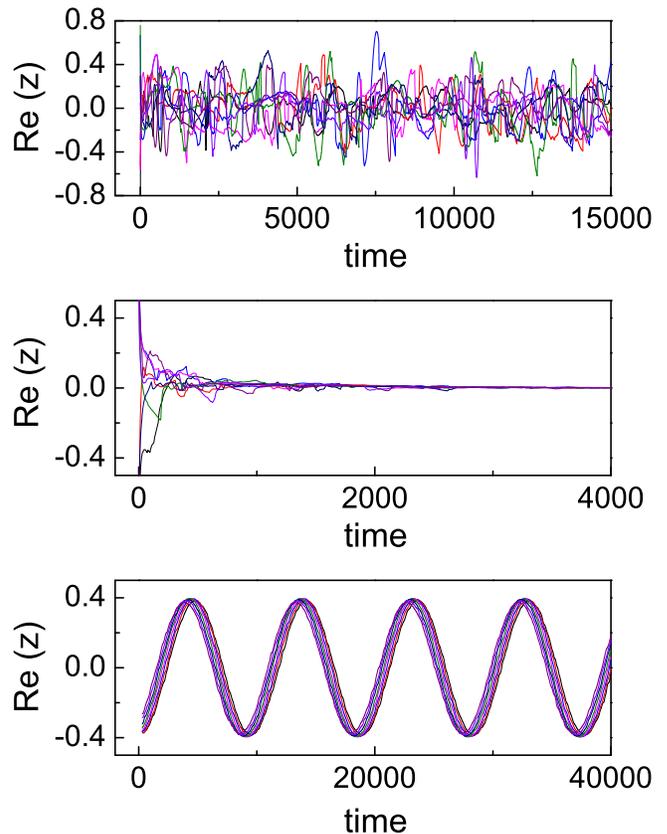}}
\caption{(Color online) Time series of the real parts of the complex
vector $z_j^\mu(t)$ for several random oscillators at $D=0.1$ (upper
panel), $D=1.0$ (middle panel) and $D=10$ (lower panel). All the
networks have fixed average network degree $\langle k\rangle =6$ and
$N =1000$. Other parameters are $\rho =10$ and $K =10$.
\label{fig1}}
\end{figure}

To quantitatively manifest the collective oscillation intensity of
the system, we choose the normalized mean \lq\lq incoherent\rq\rq\,
energy $E$ and \lq\lq coherent\rq\rq\, energy $W$ as the
characteristic variables, which are defined as
\cite{PhysRevE.62.6440,PhysRevE.68.055103}

\begin{eqnarray} \label{EqOderpara}
E &=& \left[ {\frac{{\langle \sum\nolimits_{\mu = 1}^{N}
\sum\nolimits_{j = 1}^{N_\mu}{{{\left| {{z_j^\mu}}
\right|}^2}\rangle } }}{{N_pr}}} \right], \nonumber \\
W &=& \left[{\frac{{\langle {{| {\sum\nolimits_{\mu = 1}^{N}
\sum\nolimits_{j = 1}^{N_\mu} {{z_j^\mu}} } |^2}}\rangle
}}{{{{N_p}^2}r}}} \right].
\end{eqnarray}
Here the brackets $\langle  \cdot \rangle $ denotes averaging over
time and $\left[ \cdot \right]$ stands for averaging over 20
different network realizations for each $D$ and $\rho$. A large $E$
means relatively large average oscillation amplitudes, while a
larger $W$ implicates more synchronous oscillations
\cite{PhysRevE.68.055103}. In the incoherent state, $W$ will be
nearly zero but $E$ can be large. For AD state, both $E$ and $W$
should be nearly zero. In the synchronized state, $E$ and $W$ both
have noteworthy nonzero values and they should be equal for complete
synchronization.

\begin{figure}[h]
\centerline{\includegraphics*[width=1.08 \columnwidth]{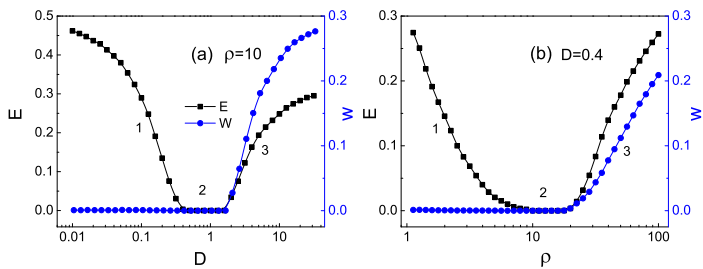}}
\caption{(Color online) The \lq\lq incoherent\rq\rq energy $E$ and
\lq\lq coherent\rq\rq energy $W$ of the limit cycles of individual
oscillators as a function of $D$ in panel (a), and $\rho$ in panel
(b). Three stages are presented as indicated by 1, 2, and 3.
\label{fig2}}
\end{figure}

The dependences of $E$ and $W$ on $D$ for fixed $\rho=10.0$ and
$\rho$ for fixed $D=0.4$ are depicted in Fig.\ref{fig2}, wherein the
curves can be divided into three stages. In stage 1, $E$ decreases
monotonously with increasing $D$($\rho$), approaching nearly zero at
a certain value of $D$($\rho$). In stage 2, both $E$ and $W$ remain
nearly zero, demonstrating the occurrence of global $AD$. Both $E$
and $W$ increase with $D$($\rho$) in stage 3, corresponding to the
synchronized state. In accordance with Fig.1, the transition from
incoherence to global AD and then to synchronization is clearly
demonstrated. It seems that increasing density plays a similar role
to increasing the mobility. Of particular interest, there exists a
notable region (stage 2) for both $D$ and $\rho$ where the system
remains stably in the AD state.

\begin{figure}[h]
\centerline{\includegraphics*[width=1.06 \columnwidth]{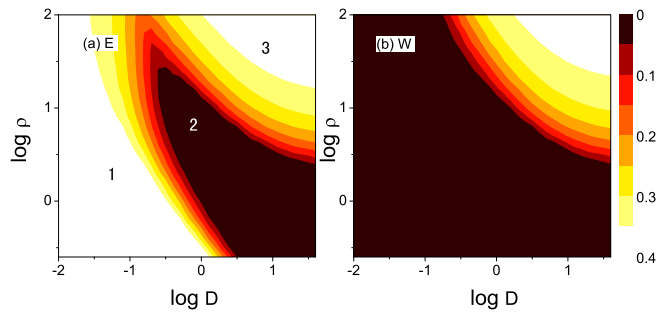}}
\caption{(Color online) Panels (a) and (b) correspond to phase
diagram $E(D, \rho)$ and $W(D, \rho)$ of the metapopulation model
respectively, both on a synthesized 1000-node BA network with
$\langle k\rangle =6$. \label{fig3}}
\end{figure}

To further get a global picture, we have perform extensive simulations to obtain the phase diagram in the
$D \sim \rho$ plane. This is shown in Fig.3, where the contour plots of $E$ and $W$ are drawn.
As expected, when both $D$ and $\rho$ are small, the system shows the
incoherent state (the light region 1 of Fig. \ref{fig3}(a)). When
$D$ and $\rho$ are increased crossing certain critical values, the system finally evolves into global AD
state (the dark region 2 of Fig.\ref{fig3}(a)). For sufficiently large $D$ and $\rho$, however, the system
may undergo a phase transition again from the death phase to
synchronized one, as shown in the light region 3 of Fig.
\ref{fig3}(a). Thus as we keep one of the parameters fixed ($D$ or
$\rho$ ) and increase the other, we observe transitions from the
incoherent (region 1) to the amplitude death (region 2) and then to
frequency synchronized state (region 3). Note that the $1\rightarrow 2$ transition is not observable with
increasing $\rho$ if the mobility rate $D$ is too small, and similarly,  the $2\rightarrow 3$ transition cannot happen if the density $\rho$ is too low.

\begin{figure}[h]
\centerline{\includegraphics*[width=1.06 \columnwidth]{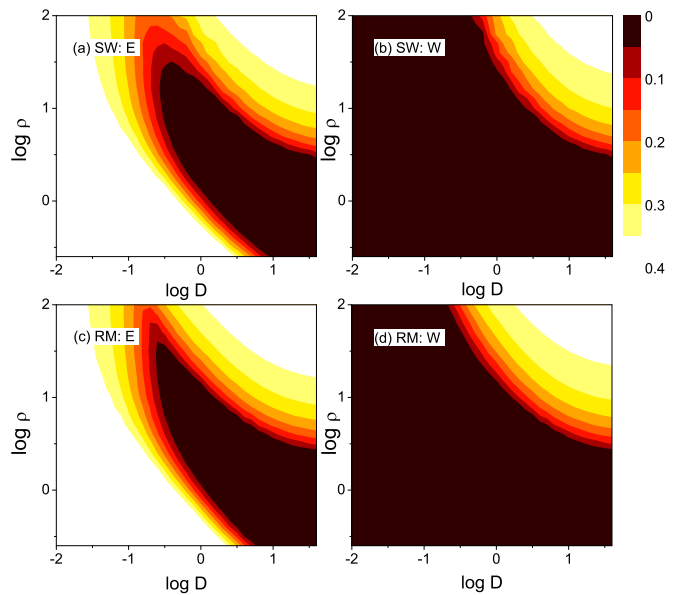}}
\caption{(Color online) Panels (a) and (b) correspond to phase
diagram $E(D, \rho)$ and $W(D, \rho)$ of the metapopulation model on
1000-node small-world networks respectively, (c) and (d) for random
networks. Other parameters are the same as Fig 3. \label{fig4}}
\end{figure}

So far, the results are all for scale-free coupled networks. One may wondering whether the interesting
findings above is sensitive to the network topology or not. Thus, we have also performed similar
studies on other types of networks, e.g., the small world
network and random network. The phase diagrams are shown in Fig.\ref{fig4} (a,b) for small-world network
and (c,d) for random network, respectively. Apparently, the qualitative behaviors are the same as those observed in
scale-free networks. The only difference is that the boundaries between different phase regions are slightly shifted.

\section{Conclusion} \label{sec7}

In summary, we have studied the collective dynamics of coupled limit cycle
oscillators defined on metapopulation networks, where different
subpopulations are connected by fluxes of individuals. By extensive
numerical simulations, we show that mobility and density play
nontrivial roles on the collective behavior of the system, by demonstrating an interesting type of
density or mobility induced transition into and out from oscillation death behavior. On one hand,
intermediate mobility rate and density can induce global oscillator
death. On the other hand, large mobility rate and density can also
 eliminate oscillator death and lead to
synchronization. In addition, we find that this nontrivial
phenomenon is robust to the network topology. The underlying mechanisms for these interesting results
are still open.  An analytical theory would surely be greatly helpful, however, it is not available at the current stage
and should certainly deserve more study.   Since
many real-life networks (cellular networks, protein networks, gene
networks, \emph{etc}.) inevitably involve variances in both the
mobility and the density, and their collective dynamics could be
modeled by coupled oscillators, these results may find a variety of
applications. Our study may also stimulate further investigation on
the emergent coherence of metapopulations of coupled oscillators.

\begin{acknowledgments}
This work was supported by the National Natural Science Foundation
of China (Grants No. 21125313, No. 20933006, No. 91027012, and No.
11205002). C.S.S. was also supported by the Key Scientific Research
Fund of Anhui Provincial Education Department (Grant No.KJ2012A189).
\end{acknowledgments}


\bibliographystyle{apsrev}
%\bibliography{OscillatorMetapopulation}

\end{document}